\PassOptionsToPackage{bookmarks=false}{hyperref}

\documentclass[sigconf]{acmart}
\copyrightyear{2020}
\acmYear{2020}
\setcopyright{acmlicensed}
\acmConference[MSR '20]{17th International Conference on Mining Software Repositories}{October 5--6, 2020}{Seoul, Republic of Korea}
\acmBooktitle{17th International Conference on Mining Software Repositories (MSR '20), October 5--6, 2020, Seoul, Republic of Korea}
\acmPrice{15.00}
\acmDOI{10.1145/3379597.3387447}
\acmISBN{978-1-4503-7517-7/20/05}

\usepackage[utf8]{inputenc}
\usepackage{url}
\usepackage{float}
\usepackage{blindtext}
\usepackage{listings,multicol}
\usepackage{graphicx}
\usepackage{balance}

\usepackage{color,soul}

\graphicspath{ {./images/} }

\begin{document}

\title{Using Large-Scale Anomaly Detection on Code to Improve Kotlin Compiler}

\author{Timofey Bryksin}
\affiliation{JetBrains Research\\
Saint Petersburg State University}
\email{t.bryksin@spbu.ru}

\author{Victor Petukhov}
\affiliation{JetBrains}
\email{victor.petukhov@jetbrains.com}

\author{Ilya Alexin}
\affiliation{Saint Petersburg State University}
\email{ilya.alexin@gmail.com}

\author{Stanislav Prikhodko}
\affiliation{Saint Petersburg State University}
\email{prikhodko.stanislav@gmail.com}

\author{Alexey Shpilman}
\affiliation{JetBrains Research\\
Higher School of Economics}
\email{alexey@shpilman.com}

\author{Vladimir Kovalenko}
\affiliation{JetBrains Research}
\email{vladimir.kovalenko@jetbrains.com}

\author{Nikita Povarov}
\affiliation{JetBrains}
\email{nikita.povarov@jetbrains.com}

\renewcommand{\shortauthors}{Bryksin et al.}

\begin{abstract}

In this work, we apply anomaly detection to source code and bytecode to facilitate the development of a programming language and its compiler. We define anomaly as a code fragment that is different from typical code written in a particular programming language. Identifying such code fragments is beneficial to both language developers and end users, since anomalies may indicate potential issues with the compiler or with runtime performance. Moreover, anomalies could correspond to problems in language design. For this study, we choose Kotlin as the target programming language. We outline and discuss approaches to obtaining vector representations of source code and bytecode and to the detection of anomalies across vectorized code snippets. The paper presents a method that aims to detect two types of anomalies: syntax tree anomalies and so-called compiler-induced anomalies that arise only in the compiled bytecode. We describe several experiments that employ different combinations of vectorization and anomaly detection techniques and discuss types of detected anomalies and their usefulness for language developers. We demonstrate that the extracted anomalies and the underlying extraction technique provide additional value for language development. 

\end{abstract}

\maketitle

\section{Introduction}

Anomaly detection techniques~\cite{hodge2004survey} have been successfully applied to a variety of practical tasks in many areas. 
These techniques help to detect cyberattacks~\cite{richardson2018anomaly, stolfo2005anomaly}, identify pathologies in medical images~\cite{baur2018deep}, and detect traces of fraudulent activities in financial data~\cite{Ahmed2016}. 

In software engineering, anomaly detection is widely applied to finding bugs~\cite{S1}, security issues~\cite{feng2003anomaly}, architectural design flaws~\cite{Oizumi2015}, workflow errors~\cite{execution-anomaly-detection}, synchronization errors in concurrent programs~\cite{Taylor1980}, and other anomalous patterns in code or other software artifacts. Definitions of an anomaly vary from study to study and are imposed by the exact task in each case.

In this study, we propose a new area of application for anomaly detection --- finding issues in programming language compilers. We define \emph{code anomalies} as fragments of code that are not typical within the community or an ecosystem of a given programming language, or machine code that is uncharacteristic within the range of output produced by the compiler. Such code could be useful to both users and developers of the language: for example, anomalous, yet actually existing, code snippets can highlight flaws in the language design or indicate problems in the performance of programs, problems in code generation, compiler optimizations, type inference, or data flow analysis, which turns them into a valuable material for compiler tests.

The task of identifying code anomalies at a scale of a language ecosystem consists in the following steps: 
    (1) retrieve large corpora of source code and compiled machine code that are representative of the whole range of conventional coding practices and compiler output, respectively; 
    (2) transform source code and machine code into a vectorized form that is digestible by anomaly detection algorithms;
    (3) run anomaly detection across vectorized data;
    (4) process the output of anomaly detection algorithms to identify, interpret, and classify meaningful anomalies.

Established industry standard languages, such as Java and C++, are not the most feasible targets for detecting anomalies at the scale of a language ecosystem. While large amounts of open source code in these languages are publicly available, mainstream compilers are very well-tested and stable, which makes it hard to identify unknown compiler issues through anomaly detection. 
At the same time, it is essential that the target language has a significant and diverse community of users.

Considering these arguments, we choose Kotlin~\cite{kotlin} and its ecosystem as a target for this study.
Since the language is relatively young, its compiler might still contain bugs and performance issues, which makes
anomaly detection more likely to provide actionable insights for the language development team.
At the same time, Kotlin is one of the most rapidly developing languages with an actively growing community and a significant ecosystem of diverse open source projects~\cite{kotlin-growth}. Finally, the development team of Kotlin is easy to reach via a public issue tracker~\cite{kotlin-yt}, which makes it easy to communicate potential findings.

The contribution of this paper is threefold:
\begin{itemize}
    \item A method for finding code anomalies for a chosen programming language that is based on vectorizing a large code corpus with code embedding techniques and then applying anomaly detection algorithms on this vector data.
    \item A set of tools implementing the proposed method as well as the dataset containing more than 4 million unique Kotlin functions collected from GitHub and the bytecode for a part of them (more than 41,000 compiled classes). 
    \item The evaluation of the proposed method on the collected dataset, which resulted in discovering several dozens of code fragments that were considered useful by the Kotlin compiler team and that were included into the Kotlin compiler test infrastructure.
    
\end{itemize}

The rest of the paper is organized as follows. In Section~\ref{related-work}, we discuss existing studies and tools that apply the anomaly detection techniques to programs. Section~\ref{overview} provides an overview of methods and techniques that are potentially applicable to finding code anomalies. Section~\ref{approach} presents our proposed approach to anomaly detection, and describes the collected dataset as well as steps of the processing pipeline. Section~\ref{evaluation} outlines our approach to evaluation of viability of the proposed method, as well as significance and practical value of the detected anomalies. Section~\ref{threats} discusses possible threats to validity of our study. Section~\ref{discussion} presents the obtained results and describes the most interesting types of discovered anomalies. Section~\ref{conclusion} concludes the paper by summarizing our results and providing possible directions for future work.

\section{Related work}\label{related-work}

Several papers and tools exist that are aiming to search for anomalies in programs. All of them introduce their own definitions of code anomalies, so their goals, methods and results also differ. 

The GrouMiner tool~\cite{S2} was developed to detect anomalous patterns in object interaction in Java programs. The approach is based on modeling object interaction with a directed acyclic graph, the nodes of which are constructor and methods calls as well as fields references, while the edges represent dependencies between them. The tool performs a static code analysis: the source code is parsed into an abstract syntax tree and an object usage graph is built. Graph-based anomaly detection methods are used to detect unusual method calls and other atypical areas of the control flow graph.

A somewhat similar idea is presented in~\cite{S3}: the authors propose mining usage models for all objects from source code as sequences of their method calls. If an abnormal usage pattern emerges in some code fragment, it is treated as an anomaly and a defect candidate. This approach is also based on static code analysis and employs graph-based anomaly detection techniques.  

Undoubtedly, object interaction anomalies are important, but they represent only one possible anomaly type, and there could be many others, for example, atypical usage of some language constructs (not involving object interaction of any kind). 

The DIDUCE tool~\cite{S1} for Java programs is based on dynamic code analysis: it runs a program and stores values for each expression found in it. It tries to induce invariants for these expressions, starting from the strictest ones and weakening them as new values are encountered. When these rules are violated, meaning that some expression gets a value that significantly differs from all previous values of this expression, this is reported as an anomaly. Papers~\cite{sekar2001fast} and~\cite{feng2003anomaly} also use dynamic code analysis but collect and analyze traces of system calls. 

Several papers~\cite{Oizumi2015, Macia2012} define code anomalies as ``code smells'' --- specific code patterns indicating possible architectural flaws. An example of such pattern is a Feature Envy smell that arises when a method interacts with other classes more than with methods and fields of its own class. Finding such code fragments in a project usually helps to improve its design.

All these approaches are helpful if one is trying to find logical errors or architectural issues in programs and therefore are targeting programming language users, not its developers. It's also worth noting that dynamic analysis algorithms don't seem like a good fit for our task since we are looking at a potentially very large code base, and running all this code does not seem feasible. Projects might have all kinds of strange dependencies, and it would require a lot of human effort to understand how all of them are supposed to be compiled and run.

In the domain of programming language development, code anomalies, thanks to being atypical code, while having been actually implemented by someone, can be used as a source of data for fuzz testing techniques, which have proven to be valuable for finding tricky issues in compilers~\cite{yang2011finding}.

Finally, in 2018 we published a preliminary discussion of the problem statement for this study~\cite{bryksin2018}.

\section{Detection of code anomalies}\label{overview}

\subsection{Anomaly detection techniques}\label{anomaly-detection}

In data mining, anomalies are defined as deviations of the observed behavior from the expected behavior and are divided into three types~\cite{Chandola}:

\begin{enumerate}
\item \textit{Point anomalies} occur when a single data instance is considered abnormal compared to other data. 
\item \textit{Contextual anomalies} occur when a single data object is anomalous in a specific context but not otherwise. 
\item \textit{Collective anomalies} occur when linked objects are observed against other objects as an anomaly. 
\end{enumerate}

In this study, we focus solely on point anomalies, but other anomaly types are also worth investigating and might lead to interesting results.

Our task implies that a dataset is unlabeled and we have no examples of code anomalies, which leads us to unsupervised anomaly detection methods~\cite{hodge2004survey}, such as Local Outlier Factor~\cite{lof} and Isolation Forest~\cite{if, if-anomaly-detection}. Several clustering methods~\cite{maimon2009introduction} could be useful for our task. Clustering methods such as DBSCAN~\cite{dbscan}, ROCK~\cite{rock}, or SNN~\cite{snn} do not require each object to be a part of a cluster, and we can see out-of-cluster objects as anomalies. One-class SVM~\cite{ocsvm} is a classification algorithm, but it assumes the presence of only one class, so the method could be used to detect outliers in unlabeled data.

Neural autoencoder~\cite{dau2014anomaly} can also serve as an anomaly detection method. Autoencoder is tasked with reconstruction of a data point through an intermediate representation. After decoder is trained on a set of datapoints, we can identify anomalous points by measuring reconstruction loss, with the assumption that the loss will be higher for outliers.

Statistical methods~\cite{prasad2013statistical} for anomaly detection measure the compliance of data to a specific probability distribution. The degree of anomalous behavior is the magnitude of the object's deviation from the distribution. For some methods in this group, an initial assumption about the distribution of data points is required, others may calculate a likely distribution that can generate the observed dataset.

\subsection{Source code embedding}

All of the anomaly detection algorithms, like most machine learning techniques, require embedding of analyzed objects in numeric vector space. To do so, first of all, we need to determine the level of structural units that will undergo analysis. It could vary from individual tokens and functions to files or even entire projects. Individual tokens and lines of code are highly dependent on their context and will not capture any significant anomalies. Functions are good candidates because most of them contain at least several lines of code that might form an anomaly and are isolated enough to represent a single operation within a class. Classes are also fitting for finding anomalies, especially in inheritance, properties use, class type parameters, etc. Using files as structural units would only allow us to detect anomalies in top-level constructions, of which there are not many and which rarely form anomalous code. Though analyzing entire projects as structural units of code may reveal some general patterns, these patterns will vary from one subject area to another and may not be informative.

Second, we need to define the level of code representation. It could be the source code itself (as text or token sets) or representations at different compilation stages. In case of Kotlin, we have access to following representations: a syntax tree; an intermediate representation for a specific compiler back-end (IR); JVM bytecode, LLVM bitcode or JavaScript code generated by the compiler. Different representations require different embedding techniques.

We can treat source code as text, which will allow the use of various natural language processing (NLP) features, such as bag-of-words or N-grams~\cite{Wang-ngrams-bug-detection, Hsiao-ngrams-js}. These features tend to capture the semantics of functions and variables names but tend to ignore the program's overall structure. NLP techniques could also be applied to bytecode since bytecode representation has a similar linear structure.

Representing a code fragment as a syntax tree introduces a wide range of embedding techniques divided into explicit and implicit approaches. Explicit approaches construct the vector from software metrics values~\cite{metrics-review, caliskan2015anonymizing} capturing lexical, syntactical, architectural and other properties. The resulting vectors are quite easy to build and comprehend: observing these values may provide a clear understanding of why this particular code fragment is considered to be anomalous. However, it is always a challenge to choose which metrics to include (for example, \cite{Varela-source-code-metrics} describes almost 300 metrics), especially when we are trying to search for anomalies of an unknown nature. Implicit approaches use features such as N-grams of token types, syntax trees hashing~\cite{jiang2007deckard, chilowicz2009syntax}, syntax trees encoding~\cite{peng2015building}, the latent vector of a neural network autoencoder~\cite{Tufano-DL-bytecode-and-AST} and other distributed code representations~\cite{code2vec-overview}. These vectors usually lack in interpretability but have proven themselves to be capable of capturing complex code properties, including semantic dependencies.

Described methods are also applicable to other available representations of Kotlin source code mentioned before: a list of LLVM bytcode instructions, a JavaScript syntax tree, and an intermediate representation (IR).

\section{The proposed method}\label{approach}
In this study, we perform a search for code anomalies in Kotlin programs that are written specifically for the JVM, since it comprises the majority of code written in Kotlin. We analyze these programs in the form of syntax trees and sequences of bytecode instructions (other representations mentioned above are available only for JavaScript and LLVM platforms). 

Our goal is to detect two types of anomalies: syntax tree anomalies and compiler-induced anomalies. A \textit{syntax tree anomaly} is a code fragment that is written in some way that is not typical for the programming language community. It could have abnormal complexity, be composed of sophisticated code constructs or in any other way differs from the rest of the code written in this language.

A \textit{compiler-induced anomaly} is a code fragment that is not an anomaly in the syntax tree form but is an anomaly in the bytecode or vice versa. We choose to call them compiler-induced because their anomalous nature is revealed only after their bytecode is obtained and analyzed. We should note, though, that this wording does not imply any negative connotation: for instance, if the compiler does a good optimizing job and turns anomalous source code into non-anomalous bytecode, this is still an unexpected behavior and, therefore, an anomaly.

\begin{figure*}[h]
\centering
\includegraphics[width=\textwidth, height=4cm]{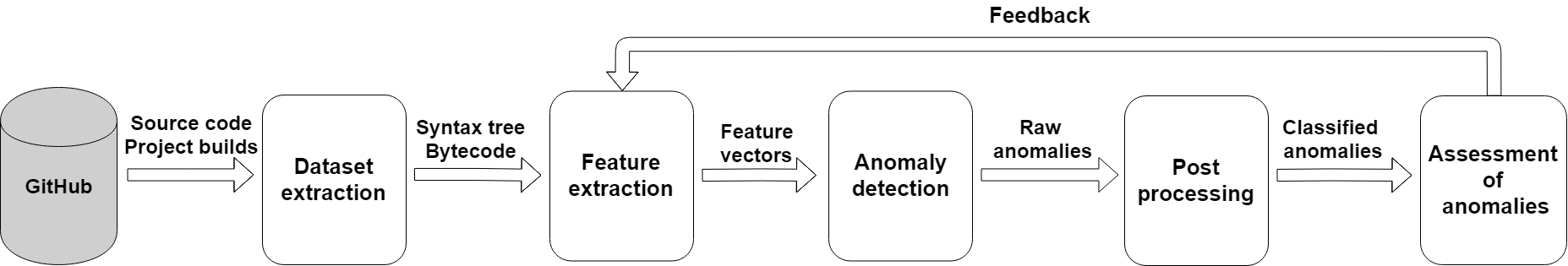}
\caption{Overview of the proposed method for extraction and assessment of anomalies} \label{pipeline}
\end{figure*}

\subsection{Dataset}
\subsubsection{Source code and syntax trees}
To collect a large dataset, we cloned GitHub repositories that were created before March 2018, stated Kotlin as their main language, and were not forks of some other projects. That resulted in 47,751 repositories containing 932,548 source files with 4,044,790 unique functions. The collected source code was transformed into syntax trees using the parser module of the Kotlin compiler. 

For the syntax trees code representation, we decided to use functions as code units. This allows us to simplify further analysis of detected anomalies, whether it is an expert assessment or a performance test. A function is an isolated unit of code, which is convenient if an anomalous fragment is to be analyzed by a compiler test because we only need to load the classes used in the function and to provide the input data it requires. Also, it will be easier for experts to analyze the function's code due to its isolation.

\subsubsection{Bytecode}
To detect compiler-induced anomalies we require both syntax tree and bytecode representations. GitHub supports publication of the project's builds packages which allows the direct collection of bytecode. Even though the dataset we collect in this way is relatively small, that kind of extraction approach is highly convenient because it allows us to collect the bytecode produced by its developers instead of figuring out the correct environment for each project and building them ourselves. We managed to obtain 41,226 compiled class files, which we then transformed into lists of JVM instructions.

Anomaly detection at functions level for bytecode representation is not as efficient as in the case of syntax trees because many Kotlin syntax constructions create additional code in bytecode representation. For example, lambdas (anonymous functions) are transformed into dummy functions. If an algorithm finds a bytecode anomaly related to such dummy function, it will be difficult to match a set of bytecode instructions to the source code. This directed us to choosing a class as a unit of evaluation for bytecode representation.

\subsection{Anomaly detection pipeline}
To detect and evaluate code anomalies, we propose the pipeline shown in Figure~\ref{pipeline}. It takes source code and project builds that are collected from GitHub as an input and converts them into syntax trees and JVM instructions lists using the dataset extraction module. Features are then extracted and anomaly detection techniques are applied. On the post-processing step, detected anomalies are classified according to their type. Finally, classes of anomalies are presented for expert evaluation. Then the pipeline's steps are adjusted for the next iteration with provided feedback.

The rest of this section discusses each step of anomaly detection in more detail. We describe assessment of individual anomalies in section \ref{sec:assessment}.

\subsubsection{Feature extraction}
We use software metrics and N-grams extraction to embed the syntax tree representation of the code. For the bytecode representation, only implicit feature extraction via N-grams is used. The bytecode has a linear structure and is rarely analyzed by humans, and fewer metrics are known for it.

Software metrics are divided into 4 groups:
\begin{itemize}
    \item general code metrics: the number of lines of code, the number of nodes and the height of the syntax tree, etc.
    \item structural metrics: nesting depth, cyclomatic complexity, number of branches in ``when'' expression, etc.
    \item external metrics of Kotlin functions: the formal arguments number, type parameters, annotations, the presence of a suspend modifier, etc.
    \item the number of particular language elements: expressions, operators, keywords, function calls, string patterns, etc.
\end{itemize}
Each function is encoded by a vector that contains the values of these metrics.

The N-grams extraction is performed as follows: the algorithm traverses a syntax tree and generates all unigrams, bigrams, ..., N-grams from connected nodes and adds a counter with value 1 for a new N-gram or increments the counter for an existing N-gram. For efficiency reasons, only parent-child relations are used to build N-grams, since considering all other relations (e.g., between sibling nodes) results in a vast number of N-grams, which will hinder the algorithm performance.

Both approaches result in a set of feature vectors. In the case of explicit representation, it is a k-dimensional vector where k is the number of metrics used. In the case of N-grams representation, it is a sparse vector where values represent counters of all N-grams met. 

\begin{figure}[h]
\centering
\includegraphics[width=1.0\columnwidth]{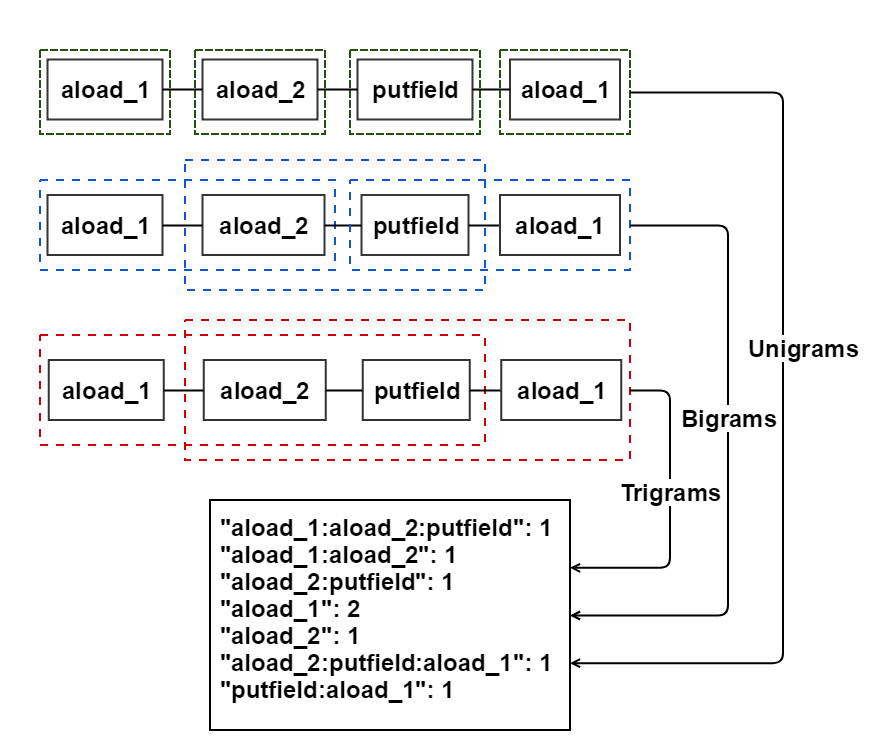}
\caption{Extraction of N-grams from sequences of bytecode instructions}\label{bytecode_to_ngrams}
\end{figure}

For compiler-induced anomalies detection, we have to process bytecode instructions sequences. N-grams are extracted incrementally from the sequence using a fixed size window. Within this window the extraction module generates all possible unigrams, bigrams, ..., N-grams, then moves the window one instruction ahead and repeats the procedure. Figure~\ref{bytecode_to_ngrams} presents an example of such extraction for $N=3$.

\subsubsection{Anomaly detection}\label{anomaly-detection-step}

In the first phase of this study, to assess general viability of the idea of anomaly detection in syntax trees, we settled on relatively simple algorithms for anomaly detection --- Local Outlier Factor~\cite{lof} and Isolation Forest~\cite{if}. We chose these algorithms because they are relatively easy to implement and trial, compared to more complex approaches.

Local Outlier Factor is based on the idea of calculating an anomaly score for each object based on its distance to k nearest neighbors in the dataset. More precisely, the local density of a data point is considered: a coefficient of the point's reachability to its k nearest neighbors. The anomaly score of a point is calculated as the ratio of local density to the mean local density in the region of a point.

\begin{figure}[h]
\centering
\includegraphics[width=0.9\columnwidth]{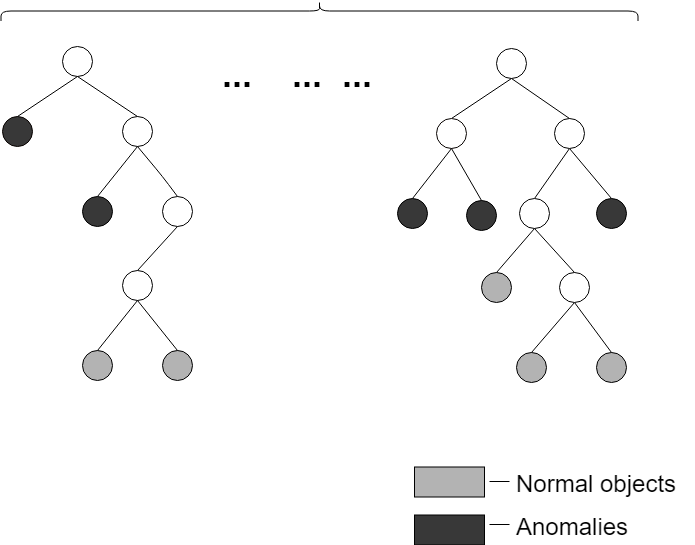}
\caption{Detecting anomalies with Isolation Forest}\label{if_architecture}
\end{figure}

Isolation Forest employs the principle of random forest to separate outlier objects from the rest of the data. Each isolating tree is constructed in the following way: at the current node, the algorithm randomly selects a feature and a value to split the data into child nodes. The process continues until all elements are separated from each other. The normality measure for an object is introduced as the average length from the root to the object's position in all isolation trees. Thus, the earlier the object is separated from the sample by isolation trees, the smaller its measure of normality will be. Figure~\ref{if_architecture} shows an example of isolation trees. Anomalous objects are marked in black, while ``normal'' objects are marked in gray.

Aforementioned algorithms are more efficient if the analyzed objects are represented by vectors of low dimensionality. When working with high-dimensional data, calculations might require too much time, hence it is necessary to fine-tune the algorithm parameters. For instance, a good choice of parameters for the Isolation Forest can sufficiently reduce processing time and memory usage without any noticeable changes in detection accuracy~\cite{if-anomaly-detection}.

A popular approach to anomaly detection in high-dimensional datasets is a neural autoencoder~\cite{autoencoder-1}, a popular type of neural networks. An autoencoder isn't so resource-intensive comparing to the Local Outlier Factor and Isolation Forest, which is crucial when we analyze the N-grams representation dataset since each feature vector contains thousands of values.

\begin{figure}[h]
\centering
\includegraphics[width=0.8\columnwidth]{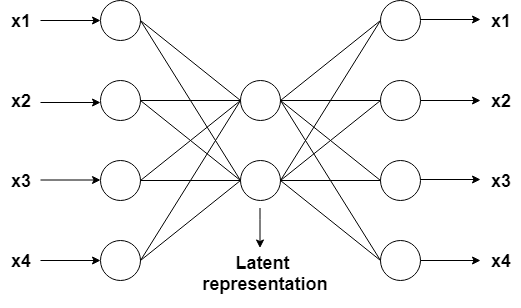}
\caption{Architecture of a neural autoencoder}\label{autencoder_architecture}
\end{figure}

Figure~\ref{autencoder_architecture} shows a basic autoencoder architecture that consists of input and output layers of the same size and a hidden layer that is significantly smaller. During the learning process, autoencoder trains to return the same vector as it receives as input. One of the results of this training is that we then have hidden layer values that comprise a vector representation (or embedding) of the input data. Another result is that if the trained autoencoder fails to reconstruct a data point, this data point may be an outlier to the dataset. Therefore, we can use the recovery error, i.e. the distance between input and output vectors as an anomaly score. 

We propose the following as a way to detect compiler-induced anomalies: first, we calculate autoencoder anomaly scores both for syntax tree and bytecode representations of the same code fragment. If these scores differ by more than a specified threshold, we consider this fragment to be a compiler-induced anomaly.

\subsubsection{Anomalies classification}\label{anomaly-classification}

We grouped anomalies into classes for further evaluation. Each anomaly was manually labeled with a set of tags. We performed classification manually to get the most reliable result possible. 

\section{Evaluation}\label{evaluation}
To assess the viability of our approach to identification of anomalies, we conduct an evaluation of significance and practical value of anomalies extracted from the dataset. In this section, we describe the process of building the evaluation set (Section \ref{sec:eval-set}), our approach to assessment of importance of individual anomalies (Section \ref{sec:assessment}), and the results of this assessment (Section \ref{sec:assess-results}).

\subsection{Building the evaluation set}\label{sec:eval-set}
In this section we describe the process of building the set of anomalies for expert assessment.
\subsubsection{Extracting anomalies from explicit source code representations}
As the first step, to build explicit representation vectors for syntax trees in the dataset, we calculated 51 software metrics for each code fragment (49 quantitative and 2 binary). We scaled the quantitative attributes to the mean of 0 and the variance of 1, and reduced dimensionality of metric vectors from 51 to 20 via Principal Component Analysis. The number of principal components was chosen manually as a compromise between the time of training the models and the persistence of the explained variance (0.8). 

We ran Local Outlier Factor and Isolation Forest outlier detection algorithms (Section~\ref{anomaly-detection-step}) against the vectorized snippets. Both of these algorithms have multiple hyperparameters. The first notable parameter is a contamination parameter, influencing the fraction of code fragments that should be marked as potential anomalies. Since the aim of the study is to find a set of anomalies that could be reviewed by experts, we chose the contamination parameter so that the classifier marked 0.01\% of the dataset, or approximately 400 out of 4 million collected functions. Specific parameter values were 0.0001 for Isolation Forest and 0.001 for Local Outlier Factor.

Other notable parameters of the outlier detection algorithms are the number of neighbors ($n\_neighbors$) for the Local Outlier Factor, which we set to 20, and the number of trees built ($n\_estimators$) for the Isolation Forest, which we set to 200. We settled on these values after several rounds of experiments and manual assessment of the resulting anomalies. As a result, these outlier detection algorithms have extracted 322 unique anomalies that we considered worthy of the attention of the Kotlin compiler team, from explicit representations of code snippets in the dataset.

\subsubsection{Extracting anomalies from implicit representations}
To build implicit representation vectors for both syntax trees and bytecode, we extracted unigrams, bigrams, and trigrams from individual instruction sets. The extraction yielded 1,708,022 unique N-grams for syntax trees and 110,835 unique N-grams for bytecode. Then we filtered rare and frequent N-grams, since they contain less information about the object, and ended up with 15982 N-grams for syntax trees and 4560 for bytecode. 

Further, we trained an autoencoder to reconstruct N-gram vectors and optimized hyperparameters of the autoencoder by a heuristic search, taking into account limitations on memory use. Resulting parameters for the autoencoder are: number of epochs --- 5, minibatch size --- 1024, compression rates --- 0.25, 0.5, and 0.75. 

To measure the \emph{anomaly score} of individual objects, we calculated the Euclidean distance between the input and the output vectors of the autoencoder. We considered anomalies all vectors of syntax trees with anomaly score exceeding 3 root mean square distances. For the compiler-induced anomalies, the threshold difference between the syntax tree and bytecode anomaly scores was set to 0.8. These values were also heuristically evaluated over multiple experiments to yield a reasonable number of anomalies. Using three autoencoder networks with different compression rates, we have extracted 191 unique syntax tree anomalies and 54 unique compiler-induced anomalies.

\subsubsection{Final processing of the evaluation set}
Using both explicit representations of source code and implicit representations of source code and bytecode, we obtained a total of 375 unique anomalies. After a quick manual filtering to remove non-interpretable items, we presented 145 remaining anomalies to Kotlin language experts for assessment of their usefulness. We describe the motivation behind expert assessment and its technique in the next section.

\subsection{Assessment of anomalies}\label{sec:assessment}
\subsubsection{Choice of assessment technique}
Code anomalies could be evaluated in several ways: by measuring the performance of the compiler on the anomalous code, by measuring runtime performance of a compiled program that contains anomalous code, or through expert assessment. 

However, there are several important concerns that make performance measurements difficult to use and interpret. 
We can only measure influence of syntax tree anomalies on the performance of the compiler's parser. It is worth noting that a parser is usually one of the simplest parts of a compiler, and performance problems in a compiler usually arise during other compilation stages: type inference, resolving, or code generation. Thus, running a performance evaluation of only a parser would be a shallow evaluation technique. At other compilation stages, performance of the compiler is highly dependent on the environment (e.g. the speed of type inference depends on the type parameters signature of the function that is called from an anomalous fragment). Moreover, a performance problem might be observed in one environment and not observed in another, and going through the complete variety of environments is an almost impossible task.

As for compiler-induced anomalies, even though we know the initial source code and the produced bytecode, it is problematic to evaluate the performance of the compiler. Data and computation that are related to a specific anomaly may be ordered differently within compiler's internal structure, which would make it very difficult to pinpoint the specific contribution of the anomalous code to the overall performance. However, it would be possible to evaluate performance of the program runtime, because we already have a corresponding bytecode to run.

A performance measurement would be relevant for assessment of the anomaly types that cause performance problems, but other anomaly types exist as well: for example, anomalies that highlight problems in the programming language design. Such problems can only be detected by an expert who deeply understands the concepts of the programming language and the internal structure of the compiler. 
\subsubsection{Expert assessment}

To cover all the anomaly types, we decided to settle on expert evaluation for assessment of the anomalies. We have invited the development team of the Kotlin compiler to serve as experts in our study. The developers of the compiler have the most comprehensive understanding of the compiler's internals, and thus can be capable of attributing individual code constructs to concrete problems. Such problems include compiler performance, program runtime performance, and language design. Two people agreed to volunteer their services as experts.

We asked the experts to evaluate each anomaly, using a five-point scale from 1 to 5. Evaluation criteria addressed three aspects: (1) whether the code fragment is typical structurally, (2) whether the code fragment can become a valuable compiler test, and (3) whether the code fragment can cause compiler performance issues. The experts were asked to select the highest rank that will hold a true statement from the following list:

\begin{itemize}
    \item \textit{Rank 1}: very typical code; could not be used as a compiler test of any kind; does not cause performance issues;
    \item \textit{Rank 2}: code with typical architecture patterns; highly unlikely to be used as a compiler test; highly unlikely to cause performance issues;
    \item \textit{Rank 3}: mostly typical code with rare use of distinctive features; could be used as a compiler test, but fragments similar to this are already used in tests; Unlikely to cause performance issues;
    \item \textit{Rank 4}: the code comprises atypical combinations of language features; could become a valuable compiler test, fragments similar to this are already used in tests;  likely to cause performance issues;
    \item \textit{Rank 5}: highly atypical code structurally; will be a unique compiler test;  most likely will cause a performance issue.
\end{itemize}

The experts evaluated anomalies together, discussing each anomaly to conclusion and providing a collective ranking decision.

We also asked the experts to rank each of the anomaly types that we obtained on the post-processing step (Section~\ref{anomaly-classification}) to assess the perceived importance of individual types.

This ranking ensures that in future experiments we are able to tune our pipeline to produce more anomalies that are of interest to the experts.

\subsection{Assessment results}\label{sec:assess-results}

\begin{figure}[h]
\includegraphics[scale=0.2]{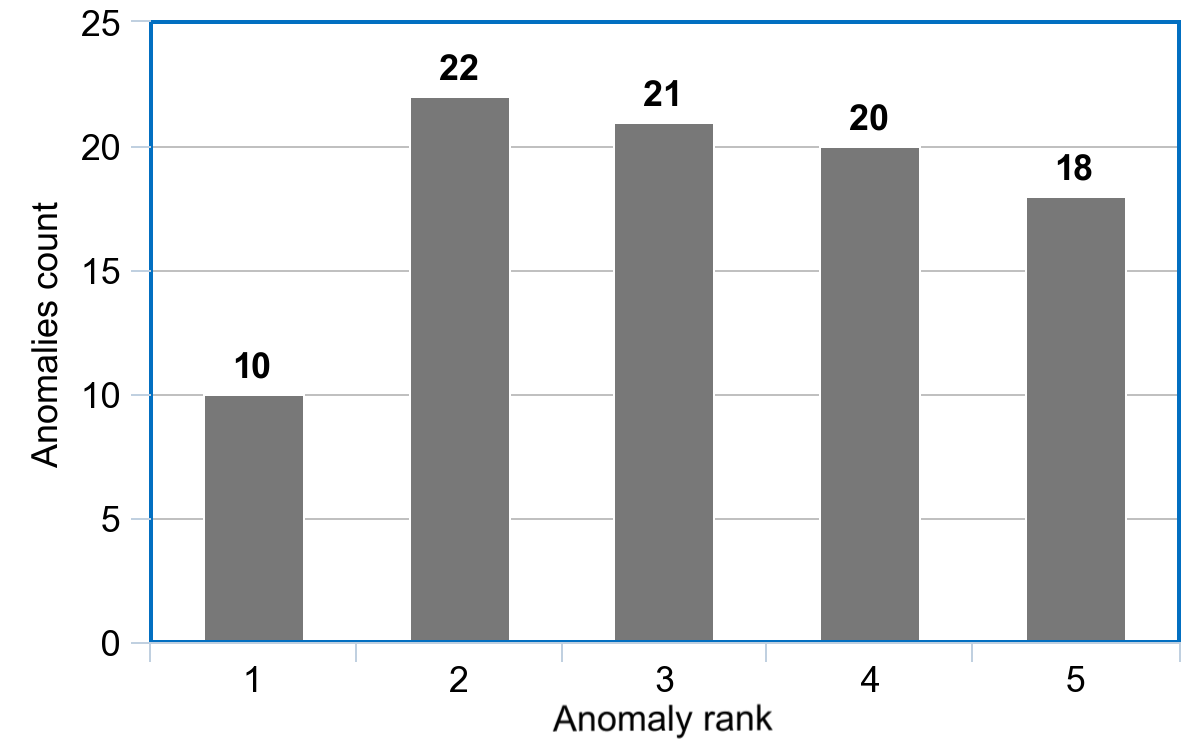}
\caption{Expert evaluation of importance for the discovered syntax tree anomalies}
\label{experts_evaluation_syntax_tree_anomalies}
\end{figure}

Figure~\ref{experts_evaluation_syntax_tree_anomalies} presents the results of expert importance assessment for syntax tree anomalies. 38 out of 91 source code anomalies selected for evaluation were considered to be important, i.e., helpful, by the experts.

Table~\ref{experts_classes_evaluation} presents assessment results for the anomaly types. The \textit{Rank} column contains the expert's evaluation rank, and the \textit{Size} column shows the number of detected anomalies of this type. 

\begin{figure}
\includegraphics[scale=0.2]{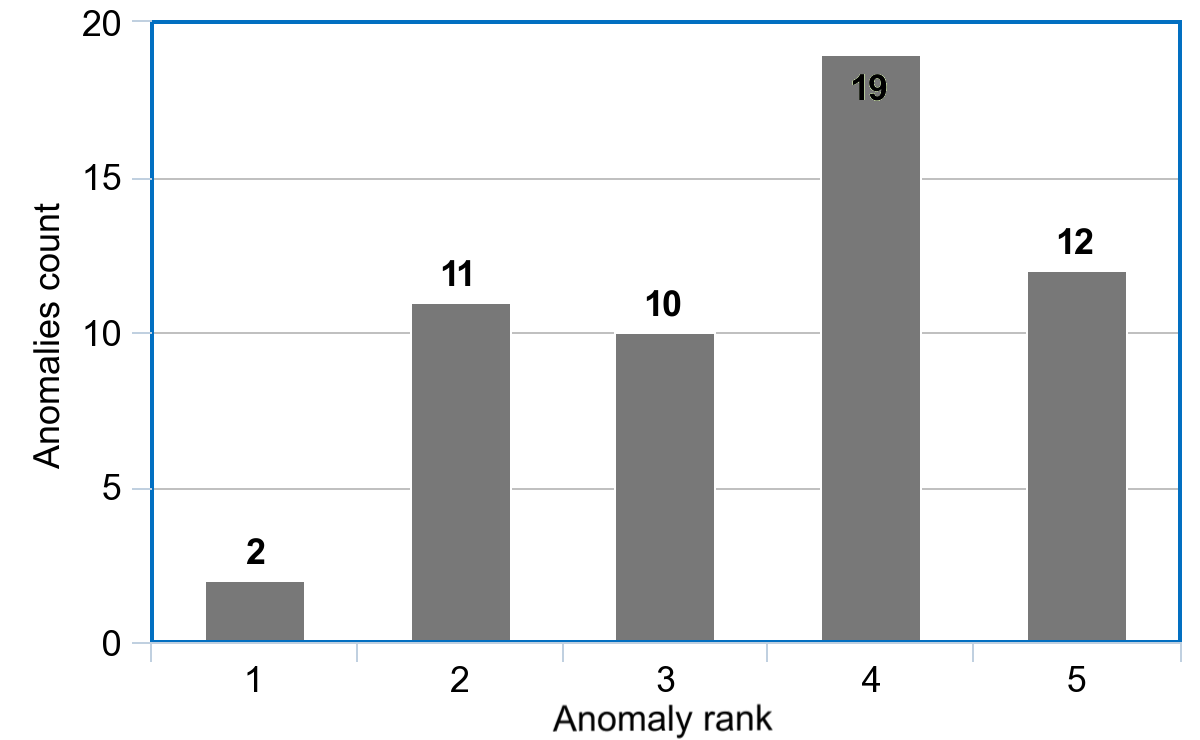}
\caption{Expert evaluation of importance for the discovered compiler-induced anomalies}
\label{experts_evaluation_conditional_anomalies}
\end{figure}

Bytecode anomalies (54 out of 145) correspond to the properties of the compiler, rather than source code. It makes such anomalies difficult to classify, hence we did not assign class labels to them. Figure~\ref{experts_evaluation_conditional_anomalies} presents the results of expert evaluations for compiler-induced anomalies. 31 out of 54 anomalies were considered helpful.

\begin{table}
\captionof{table}{Results of expert assessment of anomaly types, sorted by importance}
  \label{experts_classes_evaluation}
  \begin{tabular}{ l l l l }
    \hline
    \textbf{\#} & \textbf{Anomaly type} & \textbf{Size} & \textbf{Rank} \\ \hline
    1 & Delegates & 1 & 5 \\ 
    2 & Type arguments & 8 & 5 \\ 
    3 & ``When'' expression & 35 & 5 \\ 

    4 & Annotations & 2 & 4 \\
    5 & Call chains & 5 & 4 \\ 
    6 & Enumerations in ``when'' & 2 & 4 \\ 
    7 & ``If'' expressions & 15 & 4 \\
    8 & Nested calls & 3 & 4 \\ 
    9 & Similar call expressions & 88 & 4 \\ 
    10 & Strange code constructs & 3 & 4 \\

    11 & Assignments & 57 & 3 \\ 
    12 & Large methods & 2 & 3 \\ 
    13 & Code hierarchy & 32 & 3 \\ 
    14 & Function parameters & 17 & 3 \\ 
    15 & Multiline strings & 27 & 3 \\ 
    16 & ``Try-catch'' expressions & 1 & 3 \\ 

    17 & Arrays or maps & 32 & 2 \\ 
    18 & Class references & 2 & 2 \\ 
    19 & Concatenations & 13 & 2 \\
    20 & Lambdas & 2 & 2 \\ 
    21 & String literals & 2 & 2 \\ 
    22 & Logical expressions & 6 & 2 \\ 
    23 & Complex loops & 8 & 2 \\ 
    24 & Similar code fragments & 2 & 2 \\ 
    25 & ``Throw'' expressions & 2 & 2 \\ 

    26 & Assertions & 1 & 1 \\ 
    27 & Empty string literals & 2 & 1 \\
    28 & Local variables & 2 & 1 \\ 
    29 & Nested functions & 2 & 1 \\
    30 & Type casts & 2 & 1 \\ \hline
  \end{tabular}
\end{table}

Based on the results of expert assessment, we have compared the assessment scores for syntax tree anomalies obtained from two experiments: the first one that used explicit vectors of metrics values and Local Outlier Factor/Isolation Forest (the explicit experiment) as well as the second one that used implicit representations based on N-grams obtained with an autoencoder (the implicit experiment). The comparison shows that anomalies from both experiments were rated high, and both approaches are suitable for code anomalies detection. On the one hand, the use of explicit code metrics simplifies the analysis of the resulting anomalies, and the set of these metrics can be easily extended to search for new anomaly types. On the other hand, implicit representations of code are able to capture properties that could be rather difficult, if possible, to describe with explicit code metrics.

\subsection{Data availability}
The data that supports the findings of this study is openly available.\footnote{The dataset used in this study: \url{https://zenodo.org/record/3733794}} The created tools are also available on GitHub.\footnote{The tools supporting the proposed methods: \url{https://github.com/JetBrains-Research/kotlin-code-anomaly}}

\section{Threats to validity}\label{threats}

Our study is subject to several threats to validity and generalizability.

\textbf{External validity}. Our dataset was comprised of only open source projects available on GitHub. This limits our ability to claim that our list of discovered anomalies is exhaustive, as code of proprietary projects might contain other anomalies.

Our anomaly detection pipeline was targeted at the Kotlin ecosystem from the very start of the project. We were able to obtain our results thanks to a combination of several unique factors, notably including quick growth of the Kotlin open source community, our ability to get in contact with the compiler development team, and slightly lower maturity of the compiler, compared to other more established languages and ecosystems. Considering these unique factors, we should acknowledge that obtaining similar results for other programming languages would require building a potentially sophisticated method from scratch.

\textbf{Construct validity}. Our resulting list of anomalies is ultimately influenced by our choices regarding the set of metrics, methods for anomaly detection, and parameters of the algorithms.

\textbf{Bias of experts}. Our ranking of importance for individual anomaly types is based on estimates provided by two experts. Such quantitative estimates based on expert opinion are intrinsically subjective~\cite{kitchenham1996evaluating}. Moreover, the role of individual biases in such a small group is particularly high. Finally, our experts are the developers of Kotlin, who are deeply involved with the compiler internals; thus, their subjective perception of importance might differ from the broader group of interested public --- for example, from professional Kotlin developers.

\smallskip
We believe that these threats, while worth noting, do not invalidate our results or diminish their value. We do not claim that our methodology is universal or applicable to other languages. Instead, with this study we unravel a new scope for application of anomaly detection techniques in software engineering, and demonstrate that these techniques help to provide actionable insights to language developers.

\begin{figure*}

\includegraphics[width=\textwidth]{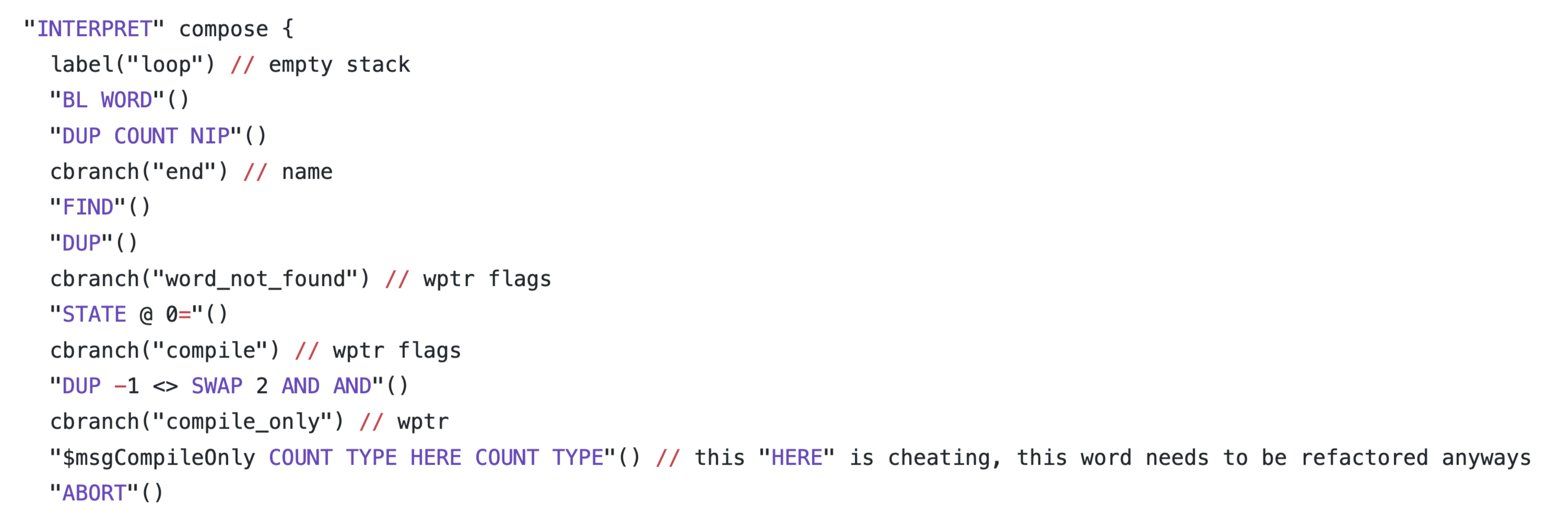}
\caption{An example of atypical, yet valid, code from a Forth implementation written in Kotlin.}
\label{anomaly_forth_compiler}
\end{figure*}

\section{Discussion}\label{discussion}
\subsection{Syntax tree anomalies}
Syntax tree anomalies are code fragments that are not typical in the Kotlin community. They are either very rare combinations of code constructs, certain combinations of code constructs repeated a large number of times, or functions with atypical characteristics, like very large object hierarchy or extensive domain-specific constructs~(Figure \ref{anomaly_forth_compiler}).

\begin{figure*}
\includegraphics[scale=0.55]{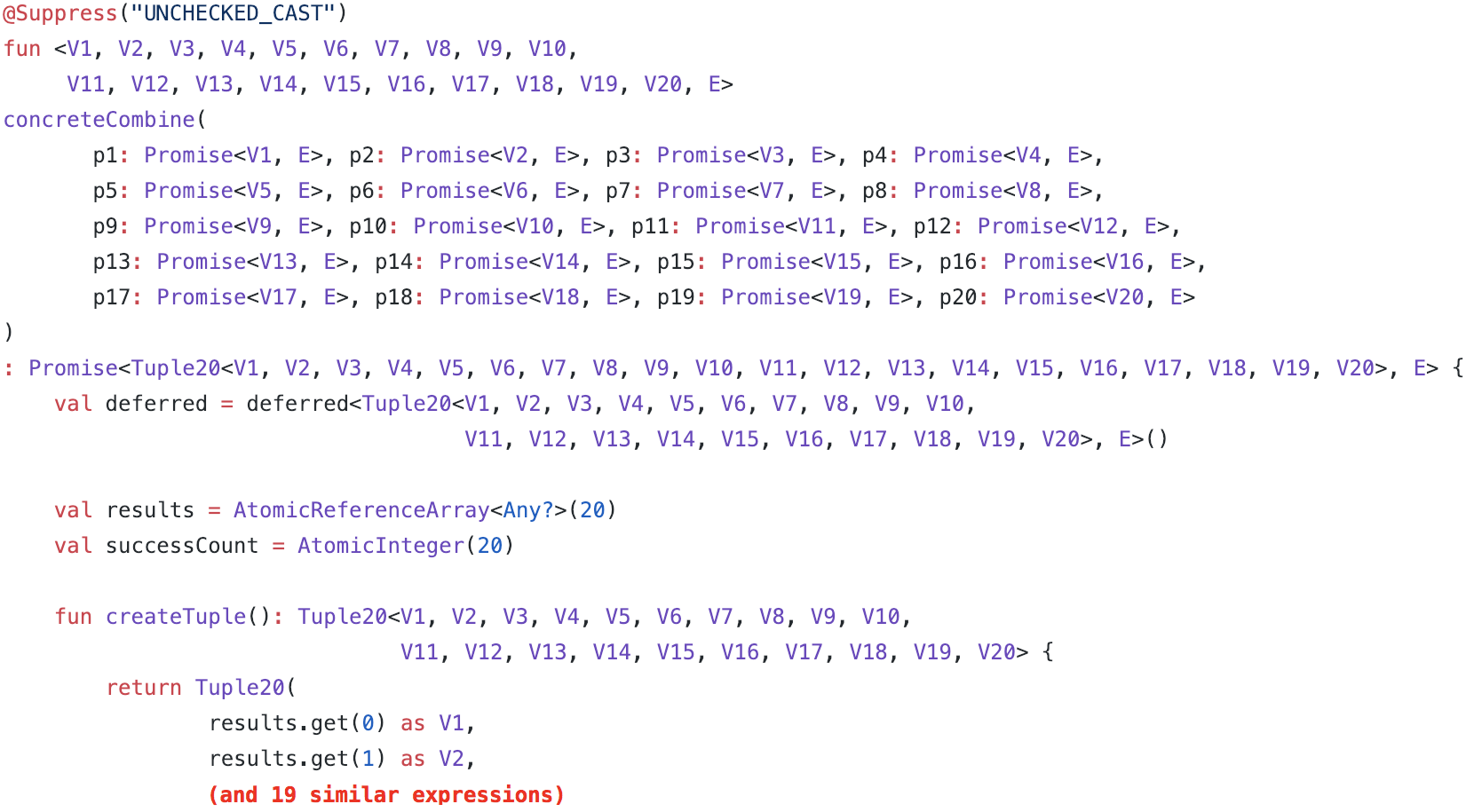}
\caption{An example of a syntax tree anomaly used as a test for performance and correctness of type inference in the compiler}
\label{anomaly_example_2_source}
\end{figure*}

The detected syntax tree anomalies were divided into three categories:
\begin{enumerate}
\item \textbf{Anomalies that correspond to language design issues}. Such anomalies were noted and archived by the Kotlin compiler team for further discussion and research. Later, such discussions could lead, for example, to introduction of a new construct to the language, deprecation of constructs use in certain cases, extension of the language standard library, refinement of the reference documentation or the language specification. Thus, the detected anomalies will be able to play a role in the development of the programming language.
\item \textbf{Anomalies that correspond to cases when some part of the compiler might work incorrectly or slowly}. The experts considered such anomalies as valuable compiler tests, and accepted to include them in the compiler's testing infrastructure. They can be used as performance tests for the compilation stages of type inference and resolving, as type inference tests for correctness, or as tests for generated bytecode. For example, the experts noted that some anomalies highlight various corner cases of the type inference as well as data flow analysis and can be very useful during refactorings in these compiler modules. An example of such an anomaly is presented in Figure~\ref{anomaly_example_2_source}. This function can act as a good test for the performance and correctness of the compiler type inference module, since processing it involves non-trivial analysis and type inference.
\item \textbf{Anomalies that correspond to potential performance issues of the program's runtime environment}. In such cases, there may also be some issues with the compiler. For example, non-optimal code generation, failures, or lack of some optimizations. Such code fragments can also be used as compiler tests; more precisely, as performance tests for the compiled programs. 
\end{enumerate}

\subsection{Compiler-induced anomalies}
Compiler-induced anomalies correspond to cases when a complex or atypical bytecode was generated by a simple or ``normal'' syntax tree (or vice versa, a non-anomalous bytecode was produced from an anomalous syntax tree).

\begin{figure*}
\includegraphics[scale=0.17]{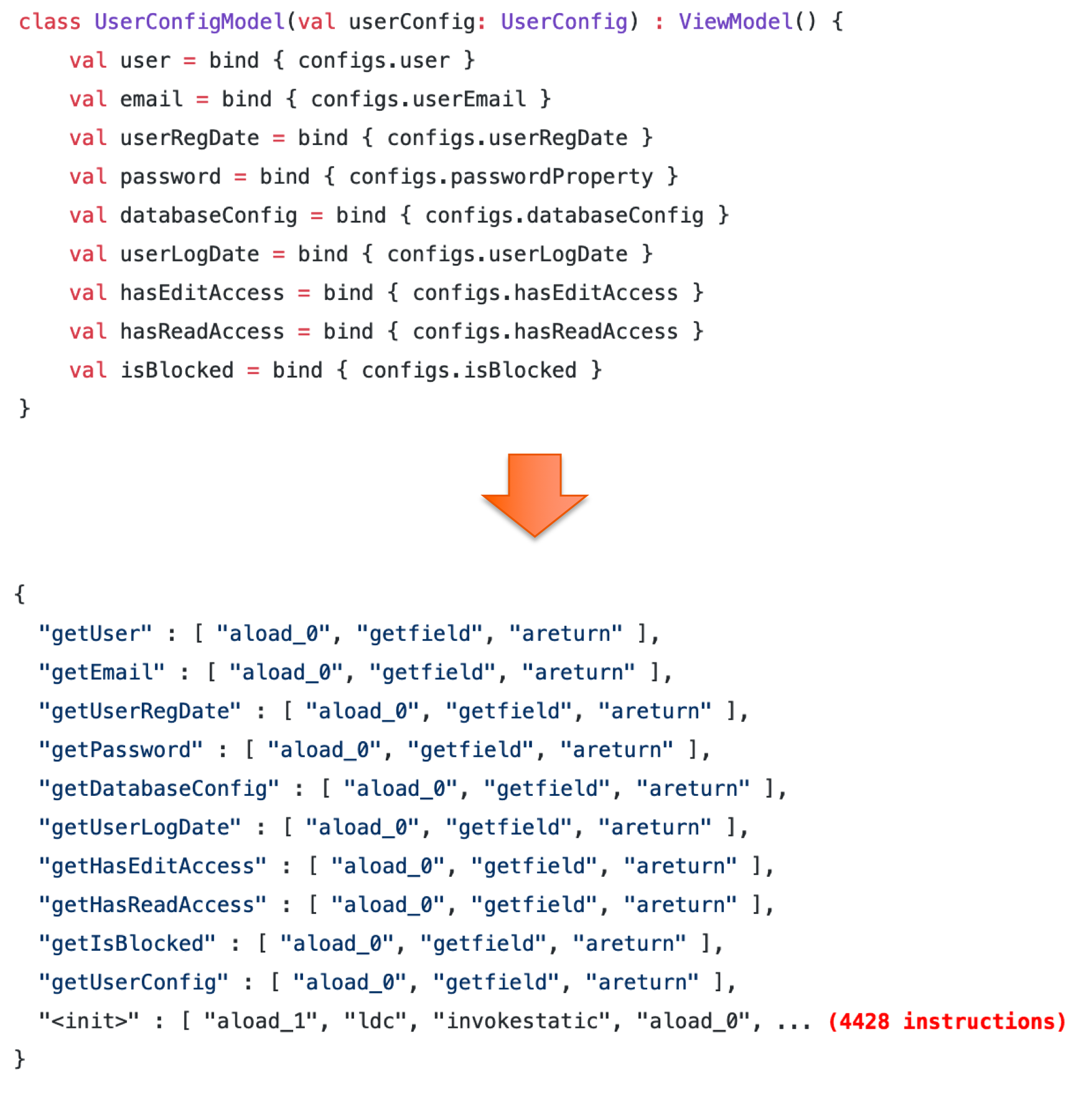}
\caption{An example of a compiler-induced anomaly highlighting an issue with function inlining}
\label{conditional_anomaly_example_1}
\end{figure*}

The detected compiler-induced anomalies were caused by the following issues:
\begin{enumerate}
\item \textbf{Non-optimal code generation cases in the compiler that manifested in some rather exotic code examples}. In such cases, the anomaly can be used as a test for bytecode generation.
\item \textbf{Rather complex functions were inlined in code fragments that were considered abnormal}. Such cases have nothing to do with compiler problems, but could be very useful for developers of libraries and frameworks, because their API is available to a potentially large number of programmers. Detection of function inlining problems could also help regular developers (i.e. users of the language) to fix performance issues in their projects. Figure~\ref{conditional_anomaly_example_1} presents an example of such an anomaly containing problematic function inlining: the framework developers wrote a large and complex function to bind properties with configuration data, which was marked ``inline''; therefore, its bytecode was copied to all places where it was called. In this example, the ``bind'' function is only called 9 times, but the bytecode is already very large and complex. This code fragment was not considered a syntax tree anomaly, but its bytecode was marked as anomalous, which helped to catch this code fragment.
\end{enumerate}
Among the suspicious code fragments there were several cases when some of the compiler optimizations worked well, and complex syntax trees were minimized to form a rather simple bytecode. Surely, such cases are not issues of any kind, but they can nevertheless be useful to compiler developers: for example, they can learn from the successful optimizations and further improve them.

\section{Conclusion}\label{conclusion}

In this paper we present two kinds of experiments to detect different types of code anomalies (syntax tree anomalies and compiler-induced anomalies), using different approaches both at the code vectorization and at the anomaly detection stages. As a result, 91 syntax tree anomalies and 54 compiler-induced anomalies were presented for assessment of importance to the Kotlin compiler developers. According to the results of the assessment, 38 syntax tree anomalies and 31 compiler-induced anomalies were considered useful (they got rank values 4 and 5). Some of these anomalies were added into the compiler testing infrastructure as performance and correctness tests for the compiler front-end and back-end; several anomalies were postponed for further discussion on possible language design issues. Based on these results and further discussions with the Kotlin compiler developers, we believe that the detected anomalies are useful and valuable for the language development, and our proposed approach proved itself successful in detection of various code anomalies. 

We outline several directions for future research:
\begin{enumerate}
\item The post-processing step of the pipeline could be automated (at least to some extent) to provide tools for classification and labeling of the found anomalies. This would allow us to focus more on finding new and more interesting types of anomalies.
\item The bytecode dataset could be increased (for example, by creating tools that automatically build projects from GitHub). At the moment, only a small part of the repositories publish builds of their projects in the ``Releases'' GitHub section. Automatic build tools would provide us with bytecode for all projects with syntactically and semantically correct code. With such dataset at hand, we could identify much more compiler-induced anomalies.
\item Compiler-induced anomalies could be used to evaluate the impact of new language features on the compilation process. This could be implemented as tests that track anomaly scores of some code examples during the compiler development process. If the anomaly scores change too much after a specific change in the compiler code, then such changes should be analyzed in detail and it should be understood what led to such an effect.
\end{enumerate}

\clearpage

\bibliographystyle{ACM-Reference-Format}
\bibliography{references.bib}


\begin{thebibliography}{42}


\ifx \showCODEN    \undefined \def \showCODEN     #1{\unskip}     \fi
\ifx \showDOI      \undefined \def \showDOI       #1{#1}\fi
\ifx \showISBNx    \undefined \def \showISBNx     #1{\unskip}     \fi
\ifx \showISBNxiii \undefined \def \showISBNxiii  #1{\unskip}     \fi
\ifx \showISSN     \undefined \def \showISSN      #1{\unskip}     \fi
\ifx \showLCCN     \undefined \def \showLCCN      #1{\unskip}     \fi
\ifx \shownote     \undefined \def \shownote      #1{#1}          \fi
\ifx \showarticletitle \undefined \def \showarticletitle #1{#1}   \fi
\ifx \showURL      \undefined \def \showURL       {\relax}        \fi
\providecommand\bibfield[2]{#2}
\providecommand\bibinfo[2]{#2}
\providecommand\natexlab[1]{#1}
\providecommand\showeprint[2][]{arXiv:#2}

\bibitem[\protect\citeauthoryear{??}{kot}{2019a}]%
        {kotlin-yt}
 \bibinfo{year}{2019}\natexlab{a}.
\newblock \bibinfo{title}{Kotlin (KT) – Bug and Issue Tracker}.
\newblock
  \bibinfo{howpublished}{\url{https://youtrack.jetbrains.com/issues/KT}}.
\newblock
\newblock
\shownote{Accessed: 2019-08-18.}


\bibitem[\protect\citeauthoryear{??}{kot}{2019b}]%
        {kotlin}
 \bibinfo{year}{2019}\natexlab{b}.
\newblock \bibinfo{title}{Kotlin Programming Language}.
\newblock \bibinfo{howpublished}{\url{https://kotlinlang.org/}}.
\newblock
\newblock
\shownote{Accessed: 2019-08-18.}


\bibitem[\protect\citeauthoryear{??}{kot}{2019c}]%
        {kotlin-growth}
 \bibinfo{year}{2019}\natexlab{c}.
\newblock \bibinfo{title}{The State of the Octoverse: top programming languages
  of 2018 – The GitHub Blog}.
\newblock
  \bibinfo{howpublished}{\url{https://github.blog/2018-11-15-state-of-the-octoverse-top-programming-languages/}}.
\newblock
\newblock
\shownote{Accessed: 2019-08-18.}


\bibitem[\protect\citeauthoryear{Ahmed, Mahmood, and Islam}{Ahmed
  et~al\mbox{.}}{2016}]%
        {Ahmed2016}
\bibfield{author}{\bibinfo{person}{Mohiuddin Ahmed},
  \bibinfo{person}{Abdun~Naser Mahmood}, {and} \bibinfo{person}{Md.~Rafiqul
  Islam}.} \bibinfo{year}{2016}\natexlab{}.
\newblock \showarticletitle{A Survey of Anomaly Detection Techniques in
  Financial Domain}.
\newblock \bibinfo{journal}{\emph{Future Gener. Comput. Syst.}}
  \bibinfo{volume}{55}, \bibinfo{number}{C} (\bibinfo{date}{Feb.}
  \bibinfo{year}{2016}), \bibinfo{pages}{278--288}.
\newblock
\showISSN{0167-739X}
\urldef\tempurl%
\url{https://doi.org/10.1016/j.future.2015.01.001}
\showDOI{\tempurl}


\bibitem[\protect\citeauthoryear{Allamanis, Barr, Devanbu, and
  Sutton}{Allamanis et~al\mbox{.}}{2017}]%
        {code2vec-overview}
\bibfield{author}{\bibinfo{person}{Miltiadis Allamanis}, \bibinfo{person}{Earl
  Barr}, \bibinfo{person}{Premkumar Devanbu}, {and} \bibinfo{person}{Charles
  Sutton}.} \bibinfo{year}{2017}\natexlab{}.
\newblock \showarticletitle{A Survey of Machine Learning for Big Code and
  Naturalness}.
\newblock \bibinfo{journal}{\emph{Comput. Surveys}}  \bibinfo{volume}{51}
  (\bibinfo{date}{09} \bibinfo{year}{2017}).
\newblock
\urldef\tempurl%
\url{https://doi.org/10.1145/3212695}
\showDOI{\tempurl}


\bibitem[\protect\citeauthoryear{Baur, Wiestler, Albarqouni, and Navab}{Baur
  et~al\mbox{.}}{2018}]%
        {baur2018deep}
\bibfield{author}{\bibinfo{person}{Christoph Baur}, \bibinfo{person}{Benedikt
  Wiestler}, \bibinfo{person}{Shadi Albarqouni}, {and} \bibinfo{person}{Nassir
  Navab}.} \bibinfo{year}{2018}\natexlab{}.
\newblock \showarticletitle{Deep autoencoding models for unsupervised anomaly
  segmentation in brain mr images}. In \bibinfo{booktitle}{\emph{International
  MICCAI Brainlesion Workshop}}. Springer, \bibinfo{pages}{161--169}.
\newblock


\bibitem[\protect\citeauthoryear{Breunig, Kriegel, Ng, and Sander}{Breunig
  et~al\mbox{.}}{2000}]%
        {lof}
\bibfield{author}{\bibinfo{person}{Markus~M. Breunig},
  \bibinfo{person}{Hans-Peter Kriegel}, \bibinfo{person}{Raymond~T. Ng}, {and}
  \bibinfo{person}{J\"{o}rg Sander}.} \bibinfo{year}{2000}\natexlab{}.
\newblock \showarticletitle{LOF: Identifying Density-based Local Outliers}.
\newblock \bibinfo{journal}{\emph{SIGMOD Rec.}} \bibinfo{volume}{29},
  \bibinfo{number}{2} (\bibinfo{date}{May} \bibinfo{year}{2000}),
  \bibinfo{pages}{93--104}.
\newblock
\showISSN{0163-5808}
\urldef\tempurl%
\url{https://doi.org/10.1145/335191.335388}
\showDOI{\tempurl}


\bibitem[\protect\citeauthoryear{Bryksin, Petukhov, Smirenko, and
  Povarov}{Bryksin et~al\mbox{.}}{2018}]%
        {bryksin2018}
\bibfield{author}{\bibinfo{person}{Timofey Bryksin}, \bibinfo{person}{Victor
  Petukhov}, \bibinfo{person}{Kirill Smirenko}, {and} \bibinfo{person}{Nikita
  Povarov}.} \bibinfo{year}{2018}\natexlab{}.
\newblock \showarticletitle{Detecting Anomalies in Kotlin Code}. In
  \bibinfo{booktitle}{\emph{Companion Proceedings for the ISSTA/ECOOP 2018
  Workshops}} (Amsterdam, Netherlands) \emph{(\bibinfo{series}{ISSTA ’18})}.
  \bibinfo{publisher}{Association for Computing Machinery},
  \bibinfo{address}{New York, NY, USA}, \bibinfo{pages}{10–12}.
\newblock
\showISBNx{9781450359399}
\urldef\tempurl%
\url{https://doi.org/10.1145/3236454.3236457}
\showDOI{\tempurl}


\bibitem[\protect\citeauthoryear{Caliskan-Islam, Harang, Liu, Narayanan, Voss,
  Yamaguchi, and Greenstadt}{Caliskan-Islam et~al\mbox{.}}{2015}]%
        {caliskan2015anonymizing}
\bibfield{author}{\bibinfo{person}{Aylin Caliskan-Islam},
  \bibinfo{person}{Richard Harang}, \bibinfo{person}{Andrew Liu},
  \bibinfo{person}{Arvind Narayanan}, \bibinfo{person}{Clare Voss},
  \bibinfo{person}{Fabian Yamaguchi}, {and} \bibinfo{person}{Rachel
  Greenstadt}.} \bibinfo{year}{2015}\natexlab{}.
\newblock \showarticletitle{De-anonymizing programmers via code stylometry}. In
  \bibinfo{booktitle}{\emph{24th $\{$USENIX$\}$ Security Symposium
  ($\{$USENIX$\}$ Security 15)}}. \bibinfo{pages}{255--270}.
\newblock


\bibitem[\protect\citeauthoryear{Chandola, Banerjee, and Kumar}{Chandola
  et~al\mbox{.}}{2009}]%
        {Chandola}
\bibfield{author}{\bibinfo{person}{Varun Chandola}, \bibinfo{person}{Arindam
  Banerjee}, {and} \bibinfo{person}{Vipin Kumar}.}
  \bibinfo{year}{2009}\natexlab{}.
\newblock \showarticletitle{Anomaly Detection: A Survey}.
\newblock \bibinfo{journal}{\emph{ACM Comput. Surv.}} \bibinfo{volume}{41},
  \bibinfo{number}{3}, Article \bibinfo{articleno}{15} (\bibinfo{date}{July}
  \bibinfo{year}{2009}), \bibinfo{numpages}{58}~pages.
\newblock
\showISSN{0360-0300}
\urldef\tempurl%
\url{https://doi.org/10.1145/1541880.1541882}
\showDOI{\tempurl}


\bibitem[\protect\citeauthoryear{Chilowicz, Duris, and Roussel}{Chilowicz
  et~al\mbox{.}}{2009}]%
        {chilowicz2009syntax}
\bibfield{author}{\bibinfo{person}{Michel Chilowicz}, \bibinfo{person}{Etienne
  Duris}, {and} \bibinfo{person}{Gilles Roussel}.}
  \bibinfo{year}{2009}\natexlab{}.
\newblock \showarticletitle{Syntax tree fingerprinting for source code
  similarity detection}. In \bibinfo{booktitle}{\emph{2009 IEEE 17th
  International Conference on Program Comprehension}}. IEEE,
  \bibinfo{pages}{243--247}.
\newblock


\bibitem[\protect\citeauthoryear{Dau, Ciesielski, and Song}{Dau
  et~al\mbox{.}}{2014}]%
        {dau2014anomaly}
\bibfield{author}{\bibinfo{person}{Hoang~Anh Dau}, \bibinfo{person}{Vic
  Ciesielski}, {and} \bibinfo{person}{Andy Song}.}
  \bibinfo{year}{2014}\natexlab{}.
\newblock \showarticletitle{Anomaly detection using replicator neural networks
  trained on examples of one class}. In \bibinfo{booktitle}{\emph{Asia-Pacific
  Conference on Simulated Evolution and Learning}}. Springer,
  \bibinfo{pages}{311--322}.
\newblock


\bibitem[\protect\citeauthoryear{Ertoz, Steinbach, and Kumar}{Ertoz
  et~al\mbox{.}}{2002}]%
        {snn}
\bibfield{author}{\bibinfo{person}{Levent Ertoz}, \bibinfo{person}{Michael
  Steinbach}, {and} \bibinfo{person}{Vipin Kumar}.}
  \bibinfo{year}{2002}\natexlab{}.
\newblock \showarticletitle{A new shared nearest neighbor clustering algorithm
  and its applications}. In \bibinfo{booktitle}{\emph{Workshop on clustering
  high dimensional data and its applications at 2nd SIAM international
  conference on data mining}}. \bibinfo{pages}{105--115}.
\newblock


\bibitem[\protect\citeauthoryear{Ester, Kriegel, Sander, Xu,
  et~al\mbox{.}}{Ester et~al\mbox{.}}{1996}]%
        {dbscan}
\bibfield{author}{\bibinfo{person}{Martin Ester}, \bibinfo{person}{Hans-Peter
  Kriegel}, \bibinfo{person}{J{\"o}rg Sander}, \bibinfo{person}{Xiaowei Xu},
  {et~al\mbox{.}}} \bibinfo{year}{1996}\natexlab{}.
\newblock \showarticletitle{A density-based algorithm for discovering clusters
  in large spatial databases with noise.}. In \bibinfo{booktitle}{\emph{Kdd}},
  Vol.~\bibinfo{volume}{96}. \bibinfo{pages}{226--231}.
\newblock


\bibitem[\protect\citeauthoryear{Feng, Kolesnikov, Fogla, Lee, and Gong}{Feng
  et~al\mbox{.}}{2003}]%
        {feng2003anomaly}
\bibfield{author}{\bibinfo{person}{Henry~Hanping Feng}, \bibinfo{person}{Oleg~M
  Kolesnikov}, \bibinfo{person}{Prahlad Fogla}, \bibinfo{person}{Wenke Lee},
  {and} \bibinfo{person}{Weibo Gong}.} \bibinfo{year}{2003}\natexlab{}.
\newblock \showarticletitle{Anomaly detection using call stack information}. In
  \bibinfo{booktitle}{\emph{Security and Privacy, 2003. Proceedings. 2003
  Symposium on}}. IEEE, \bibinfo{pages}{62--75}.
\newblock


\bibitem[\protect\citeauthoryear{Fu, Lou, Wang, and Li}{Fu
  et~al\mbox{.}}{2009}]%
        {execution-anomaly-detection}
\bibfield{author}{\bibinfo{person}{Qiang Fu}, \bibinfo{person}{Jian{-}Guang
  Lou}, \bibinfo{person}{Yi Wang}, {and} \bibinfo{person}{Jiang Li}.}
  \bibinfo{year}{2009}\natexlab{}.
\newblock \showarticletitle{Execution Anomaly Detection in Distributed Systems
  through Unstructured Log Analysis}. In \bibinfo{booktitle}{\emph{{ICDM} 2009,
  The Ninth {IEEE} International Conference on Data Mining, Miami, Florida,
  USA, 6-9 December 2009}}. \bibinfo{pages}{149--158}.
\newblock
\urldef\tempurl%
\url{https://doi.org/10.1109/ICDM.2009.60}
\showDOI{\tempurl}


\bibitem[\protect\citeauthoryear{Guha, Rastogi, and Shim}{Guha
  et~al\mbox{.}}{1999}]%
        {rock}
\bibfield{author}{\bibinfo{person}{Sudipto Guha}, \bibinfo{person}{Rajeev
  Rastogi}, {and} \bibinfo{person}{Kyuseok Shim}.}
  \bibinfo{year}{1999}\natexlab{}.
\newblock \showarticletitle{ROCK: A robust clustering algorithm for categorical
  attributes}. In \bibinfo{booktitle}{\emph{Data Engineering, 1999.
  Proceedings., 15th International Conference on}}. IEEE,
  \bibinfo{pages}{512--521}.
\newblock


\bibitem[\protect\citeauthoryear{Hangal and Lam}{Hangal and Lam}{2002}]%
        {S1}
\bibfield{author}{\bibinfo{person}{Sudheendra Hangal} {and}
  \bibinfo{person}{Monica~S. Lam}.} \bibinfo{year}{2002}\natexlab{}.
\newblock \showarticletitle{Tracking Down Software Bugs Using Automatic Anomaly
  Detection}. In \bibinfo{booktitle}{\emph{Proceedings of the 24th
  International Conference on Software Engineering}} (Orlando, Florida)
  \emph{(\bibinfo{series}{ICSE '02})}. \bibinfo{publisher}{ACM},
  \bibinfo{address}{New York, NY, USA}, \bibinfo{pages}{291--301}.
\newblock
\showISBNx{1-58113-472-X}
\urldef\tempurl%
\url{https://doi.org/10.1145/581339.581377}
\showDOI{\tempurl}


\bibitem[\protect\citeauthoryear{Hodge and Austin}{Hodge and Austin}{2004}]%
        {hodge2004survey}
\bibfield{author}{\bibinfo{person}{Victoria Hodge} {and} \bibinfo{person}{Jim
  Austin}.} \bibinfo{year}{2004}\natexlab{}.
\newblock \showarticletitle{A survey of outlier detection methodologies}.
\newblock \bibinfo{journal}{\emph{Artificial intelligence review}}
  \bibinfo{volume}{22}, \bibinfo{number}{2} (\bibinfo{year}{2004}),
  \bibinfo{pages}{85--126}.
\newblock


\bibitem[\protect\citeauthoryear{Hsiao, Cafarella, and Narayanasamy}{Hsiao
  et~al\mbox{.}}{2014}]%
        {Hsiao-ngrams-js}
\bibfield{author}{\bibinfo{person}{Chun-Hung Hsiao}, \bibinfo{person}{Michael
  Cafarella}, {and} \bibinfo{person}{Satish Narayanasamy}.}
  \bibinfo{year}{2014}\natexlab{}.
\newblock \showarticletitle{Using Web Corpus Statistics for Program Analysis}.
  In \bibinfo{booktitle}{\emph{Proceedings of the 2014 ACM International
  Conference on Object Oriented Programming Systems Languages \& Applications}}
  (Portland, Oregon, USA) \emph{(\bibinfo{series}{OOPSLA '14})}.
  \bibinfo{publisher}{ACM}, \bibinfo{address}{New York, NY, USA},
  \bibinfo{pages}{49--65}.
\newblock
\showISBNx{978-1-4503-2585-1}
\urldef\tempurl%
\url{https://doi.org/10.1145/2660193.2660226}
\showDOI{\tempurl}


\bibitem[\protect\citeauthoryear{Jiang, Misherghi, Su, and Glondu}{Jiang
  et~al\mbox{.}}{2007}]%
        {jiang2007deckard}
\bibfield{author}{\bibinfo{person}{Lingxiao Jiang}, \bibinfo{person}{Ghassan
  Misherghi}, \bibinfo{person}{Zhendong Su}, {and} \bibinfo{person}{Stephane
  Glondu}.} \bibinfo{year}{2007}\natexlab{}.
\newblock \showarticletitle{Deckard: Scalable and accurate tree-based detection
  of code clones}. In \bibinfo{booktitle}{\emph{Proceedings of the 29th
  international conference on Software Engineering}}. IEEE Computer Society,
  \bibinfo{pages}{96--105}.
\newblock


\bibitem[\protect\citeauthoryear{Kitchenham}{Kitchenham}{1996}]%
        {kitchenham1996evaluating}
\bibfield{author}{\bibinfo{person}{Barbara~Ann Kitchenham}.}
  \bibinfo{year}{1996}\natexlab{}.
\newblock \showarticletitle{Evaluating software engineering methods and tool
  part 1: The evaluation context and evaluation methods}.
\newblock \bibinfo{journal}{\emph{ACM SIGSOFT Software Engineering Notes}}
  \bibinfo{volume}{21}, \bibinfo{number}{1} (\bibinfo{year}{1996}),
  \bibinfo{pages}{11--14}.
\newblock


\bibitem[\protect\citeauthoryear{Liu, Ting, and Zhou}{Liu
  et~al\mbox{.}}{2008}]%
        {if}
\bibfield{author}{\bibinfo{person}{Fei~Tony Liu}, \bibinfo{person}{Kai~Ming
  Ting}, {and} \bibinfo{person}{Zhi-Hua Zhou}.}
  \bibinfo{year}{2008}\natexlab{}.
\newblock \showarticletitle{Isolation Forest}. In
  \bibinfo{booktitle}{\emph{Proceedings of the 2008 Eighth IEEE International
  Conference on Data Mining}} \emph{(\bibinfo{series}{ICDM '08})}.
  \bibinfo{publisher}{IEEE Computer Society}, \bibinfo{address}{Washington, DC,
  USA}, \bibinfo{pages}{413--422}.
\newblock
\showISBNx{978-0-7695-3502-9}
\urldef\tempurl%
\url{https://doi.org/10.1109/ICDM.2008.17}
\showDOI{\tempurl}


\bibitem[\protect\citeauthoryear{Liu, Ting, and Zhou}{Liu
  et~al\mbox{.}}{2012}]%
        {if-anomaly-detection}
\bibfield{author}{\bibinfo{person}{Fei~Tony Liu}, \bibinfo{person}{Kai~Ming
  Ting}, {and} \bibinfo{person}{Zhi-Hua Zhou}.}
  \bibinfo{year}{2012}\natexlab{}.
\newblock \showarticletitle{Isolation-Based Anomaly Detection}.
\newblock \bibinfo{journal}{\emph{ACM Trans. Knowl. Discov. Data}}
  \bibinfo{volume}{6}, \bibinfo{number}{1}, Article \bibinfo{articleno}{3}
  (\bibinfo{date}{March} \bibinfo{year}{2012}), \bibinfo{numpages}{39}~pages.
\newblock
\showISSN{1556-4681}
\urldef\tempurl%
\url{https://doi.org/10.1145/2133360.2133363}
\showDOI{\tempurl}


\bibitem[\protect\citeauthoryear{{Macia}, {Arcoverde}, {Garcia}, {Chavez}, and
  {von Staa}}{{Macia} et~al\mbox{.}}{2012}]%
        {Macia2012}
\bibfield{author}{\bibinfo{person}{I. {Macia}}, \bibinfo{person}{R.
  {Arcoverde}}, \bibinfo{person}{A. {Garcia}}, \bibinfo{person}{C. {Chavez}},
  {and} \bibinfo{person}{A. {von Staa}}.} \bibinfo{year}{2012}\natexlab{}.
\newblock \showarticletitle{On the Relevance of Code Anomalies for Identifying
  Architecture Degradation Symptoms}. In \bibinfo{booktitle}{\emph{2012 16th
  European Conference on Software Maintenance and Reengineering}}.
  \bibinfo{pages}{277--286}.
\newblock
\showISSN{1534-5351}
\urldef\tempurl%
\url{https://doi.org/10.1109/CSMR.2012.35}
\showDOI{\tempurl}


\bibitem[\protect\citeauthoryear{Maimon and Rokach}{Maimon and Rokach}{2009}]%
        {maimon2009introduction}
\bibfield{author}{\bibinfo{person}{Oded Maimon} {and} \bibinfo{person}{Lior
  Rokach}.} \bibinfo{year}{2009}\natexlab{}.
\newblock \showarticletitle{Introduction to knowledge discovery and data
  mining}.
\newblock In \bibinfo{booktitle}{\emph{Data Mining and Knowledge Discovery
  Handbook}}. \bibinfo{publisher}{Springer}, \bibinfo{pages}{1--15}.
\newblock


\bibitem[\protect\citeauthoryear{Nguyen, Nguyen, Pham, Al-Kofahi, and
  Nguyen}{Nguyen et~al\mbox{.}}{2009}]%
        {S2}
\bibfield{author}{\bibinfo{person}{Tung~Thanh Nguyen},
  \bibinfo{person}{Hoan~Anh Nguyen}, \bibinfo{person}{Nam~H. Pham},
  \bibinfo{person}{Jafar~M. Al-Kofahi}, {and} \bibinfo{person}{Tien~N.
  Nguyen}.} \bibinfo{year}{2009}\natexlab{}.
\newblock \showarticletitle{Graph-based Mining of Multiple Object Usage
  Patterns}. In \bibinfo{booktitle}{\emph{Proceedings of the the 7th Joint
  Meeting of the European Software Engineering Conference and the ACM SIGSOFT
  Symposium on The Foundations of Software Engineering}} (Amsterdam, The
  Netherlands) \emph{(\bibinfo{series}{ESEC/FSE '09})}.
  \bibinfo{publisher}{ACM}, \bibinfo{address}{New York, NY, USA},
  \bibinfo{pages}{383--392}.
\newblock
\showISBNx{978-1-60558-001-2}
\urldef\tempurl%
\url{https://doi.org/10.1145/1595696.1595767}
\showDOI{\tempurl}


\bibitem[\protect\citeauthoryear{Nuez-Varela, Prez-Gonzalez, Martnez-Perez, and
  Soubervielle-Montalvo}{Nuez-Varela et~al\mbox{.}}{2017}]%
        {metrics-review}
\bibfield{author}{\bibinfo{person}{Alberto~S. Nuez-Varela},
  \bibinfo{person}{Hctor~G. Prez-Gonzalez}, \bibinfo{person}{Francisco~E.
  Martnez-Perez}, {and} \bibinfo{person}{Carlos Soubervielle-Montalvo}.}
  \bibinfo{year}{2017}\natexlab{}.
\newblock \showarticletitle{Source Code Metrics}.
\newblock \bibinfo{journal}{\emph{J. Syst. Softw.}} \bibinfo{volume}{128},
  \bibinfo{number}{C} (\bibinfo{date}{June} \bibinfo{year}{2017}),
  \bibinfo{pages}{164--197}.
\newblock
\showISSN{0164-1212}
\urldef\tempurl%
\url{https://doi.org/10.1016/j.jss.2017.03.044}
\showDOI{\tempurl}


\bibitem[\protect\citeauthoryear{Oizumi, Garcia, Colanzi, Ferreira, and
  Staa}{Oizumi et~al\mbox{.}}{2015}]%
        {Oizumi2015}
\bibfield{author}{\bibinfo{person}{Willian~N. Oizumi},
  \bibinfo{person}{Alessandro~F. Garcia}, \bibinfo{person}{Thelma~E. Colanzi},
  \bibinfo{person}{Manuele Ferreira}, {and} \bibinfo{person}{Arndt~V. Staa}.}
  \bibinfo{year}{2015}\natexlab{}.
\newblock \showarticletitle{On the relationship of code-anomaly agglomerations
  and architectural problems}.
\newblock \bibinfo{journal}{\emph{Journal of Software Engineering Research and
  Development}} \bibinfo{volume}{3}, \bibinfo{number}{1} (\bibinfo{date}{10
  Jul} \bibinfo{year}{2015}), \bibinfo{pages}{11}.
\newblock
\showISSN{2195-1721}
\urldef\tempurl%
\url{https://doi.org/10.1186/s40411-015-0025-y}
\showDOI{\tempurl}


\bibitem[\protect\citeauthoryear{Peng, Mou, Li, Liu, Zhang, and Jin}{Peng
  et~al\mbox{.}}{2015}]%
        {peng2015building}
\bibfield{author}{\bibinfo{person}{Hao Peng}, \bibinfo{person}{Lili Mou},
  \bibinfo{person}{Ge Li}, \bibinfo{person}{Yuxuan Liu}, \bibinfo{person}{Lu
  Zhang}, {and} \bibinfo{person}{Zhi Jin}.} \bibinfo{year}{2015}\natexlab{}.
\newblock \showarticletitle{Building program vector representations for deep
  learning}. In \bibinfo{booktitle}{\emph{International Conference on Knowledge
  Science, Engineering and Management}}. Springer, \bibinfo{pages}{547--553}.
\newblock


\bibitem[\protect\citeauthoryear{Prasad and Krishna}{Prasad and
  Krishna}{2013}]%
        {prasad2013statistical}
\bibfield{author}{\bibinfo{person}{YA~Siva Prasad} {and}
  \bibinfo{person}{G~Rama Krishna}.} \bibinfo{year}{2013}\natexlab{}.
\newblock \showarticletitle{Statistical Anomaly Detection Technique for Real
  Time Datasets}.
\newblock \bibinfo{journal}{\emph{International Journal of Computer Trends and
  Technology (IJCTT)--volume}}  \bibinfo{volume}{6} (\bibinfo{year}{2013}).
\newblock


\bibitem[\protect\citeauthoryear{Randy C.~Paffenroth}{Randy
  C.~Paffenroth}{2017}]%
        {autoencoder-1}
\bibfield{author}{\bibinfo{person}{Chong~Zhou Randy C.~Paffenroth}.}
  \bibinfo{year}{2017}\natexlab{}.
\newblock \showarticletitle{Anomaly Detection with Robust Deep Autoencoders}.
  In \bibinfo{booktitle}{\emph{KDD '17 Proceedings of the 23rd ACM SIGKDD
  International Conference on Knowledge Discovery and Data Mining}}
  \emph{(\bibinfo{series}{KDD '17})}. \bibinfo{publisher}{ACM},
  \bibinfo{address}{New York, NY, USA}, \bibinfo{pages}{665--674}.
\newblock
\showISBNx{978-1-4503-4887-4}
\urldef\tempurl%
\url{https://doi.org/10.1145/3097983.3098052}
\showDOI{\tempurl}


\bibitem[\protect\citeauthoryear{Richardson, Radford, Davis, Hines, and
  Pekarek}{Richardson et~al\mbox{.}}{2018}]%
        {richardson2018anomaly}
\bibfield{author}{\bibinfo{person}{Bartley~D Richardson},
  \bibinfo{person}{Benjamin~J Radford}, \bibinfo{person}{Shawn~E Davis},
  \bibinfo{person}{Keegan Hines}, {and} \bibinfo{person}{David Pekarek}.}
  \bibinfo{year}{2018}\natexlab{}.
\newblock \showarticletitle{Anomaly Detection in Cyber Network Data Using a
  Cyber Language Approach}.
\newblock \bibinfo{journal}{\emph{arXiv preprint arXiv:1808.10742}}
  (\bibinfo{year}{2018}).
\newblock


\bibitem[\protect\citeauthoryear{Sch\"{o}lkopf, Williamson, Smola,
  Shawe-Taylor, and Platt}{Sch\"{o}lkopf et~al\mbox{.}}{1999}]%
        {ocsvm}
\bibfield{author}{\bibinfo{person}{Bernhard Sch\"{o}lkopf},
  \bibinfo{person}{Robert Williamson}, \bibinfo{person}{Alex Smola},
  \bibinfo{person}{John Shawe-Taylor}, {and} \bibinfo{person}{John Platt}.}
  \bibinfo{year}{1999}\natexlab{}.
\newblock \showarticletitle{Support Vector Method for Novelty Detection}. In
  \bibinfo{booktitle}{\emph{Proceedings of the 12th International Conference on
  Neural Information Processing Systems}} (Denver, CO)
  \emph{(\bibinfo{series}{NIPS'99})}. \bibinfo{publisher}{MIT Press},
  \bibinfo{address}{Cambridge, MA, USA}, \bibinfo{pages}{582--588}.
\newblock
\urldef\tempurl%
\url{http://dl.acm.org/citation.cfm?id=3009657.3009740}
\showURL{%
\tempurl}


\bibitem[\protect\citeauthoryear{Sekar, Bendre, Dhurjati, and Bollineni}{Sekar
  et~al\mbox{.}}{2001}]%
        {sekar2001fast}
\bibfield{author}{\bibinfo{person}{R Sekar}, \bibinfo{person}{Mugdha Bendre},
  \bibinfo{person}{Dinakar Dhurjati}, {and} \bibinfo{person}{Pradeep
  Bollineni}.} \bibinfo{year}{2001}\natexlab{}.
\newblock \showarticletitle{A fast automaton-based method for detecting
  anomalous program behaviors}. In \bibinfo{booktitle}{\emph{sp}}. IEEE,
  \bibinfo{pages}{0144}.
\newblock


\bibitem[\protect\citeauthoryear{Stolfo, Hershkop, Bui, Ferster, and
  Wang}{Stolfo et~al\mbox{.}}{2005}]%
        {stolfo2005anomaly}
\bibfield{author}{\bibinfo{person}{Salvatore~J Stolfo}, \bibinfo{person}{Shlomo
  Hershkop}, \bibinfo{person}{Linh~H Bui}, \bibinfo{person}{Ryan Ferster},
  {and} \bibinfo{person}{Ke Wang}.} \bibinfo{year}{2005}\natexlab{}.
\newblock \showarticletitle{Anomaly detection in computer security and an
  application to file system accesses}. In
  \bibinfo{booktitle}{\emph{International Symposium on Methodologies for
  Intelligent Systems}}. Springer, \bibinfo{pages}{14--28}.
\newblock


\bibitem[\protect\citeauthoryear{{Taylor} and {Osterweil}}{{Taylor} and
  {Osterweil}}{1980}]%
        {Taylor1980}
\bibfield{author}{\bibinfo{person}{R.~N. {Taylor}} {and} \bibinfo{person}{L.~J.
  {Osterweil}}.} \bibinfo{year}{1980}\natexlab{}.
\newblock \showarticletitle{Anomaly Detection in Concurrent Software by Static
  Data Flow Analysis}.
\newblock \bibinfo{journal}{\emph{IEEE Transactions on Software Engineering}}
  \bibinfo{volume}{SE-6}, \bibinfo{number}{3} (\bibinfo{date}{May}
  \bibinfo{year}{1980}), \bibinfo{pages}{265--278}.
\newblock
\showISSN{0098-5589}
\urldef\tempurl%
\url{https://doi.org/10.1109/TSE.1980.234488}
\showDOI{\tempurl}


\bibitem[\protect\citeauthoryear{Tufano, Watson, Bavota, Di~Penta, White, and
  Poshyvanyk}{Tufano et~al\mbox{.}}{2018}]%
        {Tufano-DL-bytecode-and-AST}
\bibfield{author}{\bibinfo{person}{Michele Tufano}, \bibinfo{person}{Cody
  Watson}, \bibinfo{person}{Gabriele Bavota}, \bibinfo{person}{Massimiliano
  Di~Penta}, \bibinfo{person}{Martin White}, {and} \bibinfo{person}{Denys
  Poshyvanyk}.} \bibinfo{year}{2018}\natexlab{}.
\newblock \showarticletitle{Deep Learning Similarities from Different
  Representations of Source Code}. In \bibinfo{booktitle}{\emph{Proceedings of
  the 15th International Conference on Mining Software Repositories}}
  (Gothenburg, Sweden) \emph{(\bibinfo{series}{MSR '18})}.
  \bibinfo{publisher}{ACM}, \bibinfo{address}{New York, NY, USA},
  \bibinfo{pages}{542--553}.
\newblock
\showISBNx{978-1-4503-5716-6}
\urldef\tempurl%
\url{https://doi.org/10.1145/3196398.3196431}
\showDOI{\tempurl}


\bibitem[\protect\citeauthoryear{Varela, Perez-Gonzalez, Martinez, and
  Soubervielle-Montalvo}{Varela et~al\mbox{.}}{2017}]%
        {Varela-source-code-metrics}
\bibfield{author}{\bibinfo{person}{Alberto Varela}, \bibinfo{person}{Hector
  Perez-Gonzalez}, \bibinfo{person}{Francisco Martinez}, {and}
  \bibinfo{person}{Carlos Soubervielle-Montalvo}.}
  \bibinfo{year}{2017}\natexlab{}.
\newblock \showarticletitle{Source Code Metrics: A Systematic Mapping Study}.
\newblock \bibinfo{journal}{\emph{Journal of Systems and Software}}
  \bibinfo{volume}{128} (\bibinfo{date}{04} \bibinfo{year}{2017}).
\newblock
\urldef\tempurl%
\url{https://doi.org/10.1016/j.jss.2017.03.044}
\showDOI{\tempurl}


\bibitem[\protect\citeauthoryear{Wang, Chollak, Movshovitz-Attias, and
  Tan}{Wang et~al\mbox{.}}{2016}]%
        {Wang-ngrams-bug-detection}
\bibfield{author}{\bibinfo{person}{Song Wang}, \bibinfo{person}{Devin Chollak},
  \bibinfo{person}{Dana Movshovitz-Attias}, {and} \bibinfo{person}{Lin Tan}.}
  \bibinfo{year}{2016}\natexlab{}.
\newblock \showarticletitle{Bugram: Bug Detection with N-gram Language Models}.
  In \bibinfo{booktitle}{\emph{Proceedings of the 31st IEEE/ACM International
  Conference on Automated Software Engineering}} (Singapore, Singapore)
  \emph{(\bibinfo{series}{ASE 2016})}. \bibinfo{publisher}{ACM},
  \bibinfo{address}{New York, NY, USA}, \bibinfo{pages}{708--719}.
\newblock
\showISBNx{978-1-4503-3845-5}
\urldef\tempurl%
\url{https://doi.org/10.1145/2970276.2970341}
\showDOI{\tempurl}


\bibitem[\protect\citeauthoryear{Wasylkowski, Zeller, and Lindig}{Wasylkowski
  et~al\mbox{.}}{2007}]%
        {S3}
\bibfield{author}{\bibinfo{person}{Andrzej Wasylkowski},
  \bibinfo{person}{Andreas Zeller}, {and} \bibinfo{person}{Christian Lindig}.}
  \bibinfo{year}{2007}\natexlab{}.
\newblock \showarticletitle{Detecting Object Usage Anomalies}. In
  \bibinfo{booktitle}{\emph{Proceedings of the the 6th Joint Meeting of the
  European Software Engineering Conference and the ACM SIGSOFT Symposium on The
  Foundations of Software Engineering}} (Dubrovnik, Croatia)
  \emph{(\bibinfo{series}{ESEC-FSE '07})}. \bibinfo{publisher}{ACM},
  \bibinfo{address}{New York, NY, USA}, \bibinfo{pages}{35--44}.
\newblock
\showISBNx{978-1-59593-811-4}
\urldef\tempurl%
\url{https://doi.org/10.1145/1287624.1287632}
\showDOI{\tempurl}


\bibitem[\protect\citeauthoryear{Yang, Chen, Eide, and Regehr}{Yang
  et~al\mbox{.}}{2011}]%
        {yang2011finding}
\bibfield{author}{\bibinfo{person}{Xuejun Yang}, \bibinfo{person}{Yang Chen},
  \bibinfo{person}{Eric Eide}, {and} \bibinfo{person}{John Regehr}.}
  \bibinfo{year}{2011}\natexlab{}.
\newblock \showarticletitle{Finding and understanding bugs in C compilers}. In
  \bibinfo{booktitle}{\emph{ACM SIGPLAN Notices}}, Vol.~\bibinfo{volume}{46}.
  ACM, \bibinfo{pages}{283--294}.
\newblock


\end{thebibliography}

\end{document}